\documentclass[12pt]{iopart}
\usepackage{iopams}
\usepackage{graphicx}

\newcommand{\APJ}{{\it Astrophys. J.\/} }
\newcommand{\MNRAS}{{\it Mon. Not. R. Aston. Soc.\/} }
\newcommand{\ARAA}{{\it Ann.\ Rev.\ Astron.\ Astrophys.\/} }

\begin{document}
\title{Gravitational Lensing Constraints on Dynamical and Coupled Dark
Energy}

\author{G La Vacca$^{1,3}$ and L P L Colombo$^{2,3,4}$}
\address{$^1$ Dipartimento di Fisica Teorica e Nucleare, Universit\`a di
Pavia, via A. Bassi, 6 I-27100 Pavia, Italy}
\address{$^2$ Dipartimento di Fisica ``G Occhialini'', Universit\`a di
Milano-Bicocca, Piazza della Scienza, 3 I-20126 Milano, Italy} 
\address{$^3$ INFN Sezione di Milano-Bicocca} 
\address{$^4$ Department of Physics \& Astronomy, 
University of Southern California, Los Angeles, CA 90089-0484 }
\ead{Giuseppe.Lavacca@mib.infn.it}

\begin{abstract}
Upcoming Weak Lensing (WL) surveys can be used to constrain Dark
Energy (DE) properties, namely if tomographic techniques are used to
improve their sensitivity. In this work, we use a Fisher matrix
technique to compare the power of CMB anisotropy and polarization data
with tomographic WL data, in constraining DE parameters. Adding WL
data to available CMB data improves the detection of all cosmological
parameters, but the impact is really strong when DE--DM coupling is
considered, as WL tomography can then succeed to reduce the errors on
some parameters by factors $>10~.$
\end{abstract}



\section{Introduction}
\label{sec:Intro}

The first data system requiring Dark Energy (DE) concerned cosmic
acceleration, detected through high-redshift
supernovae~\cite{sn}. CMB~\cite{cmb} and deep sample~\cite{lss} data
supported also the DE case, showing that the density parameter for
non--relativistic matter $\Omega_{0,m} \sim 0.3$, while the total
density parameter $\Omega_0 \sim 1~.$

In the most popular scenario, DE is ascribed to a cosmological
constant $\Lambda$. Alternative options include a self--interacting
scalar field, $\phi$ (quintessence or dynamical
DE~\cite{quintessence,sugra}) and modifications of General
Relativity~\cite{gr}.

It is known that models with $\Lambda$ ($\Lambda $CDM) apparently
accommodate all available data systems. The problem is the physical
origin of $\Lambda,$ which can be a false vacuum; this however causes
well known {\it fine tuning} and {\it coincidence} problems.

The former problem is partially eased in dynamical DE (dDE) scenarios,
when self interaction is due to a {\it tracking} potential
$V(\phi)$~\cite{tracking}. If $V(\phi)$ is SUGRA \cite{sugra} , the
fit with data is at least as good as for $\Lambda$CDM
\cite{lambdalim}.

In the attempt to ease the coincidence problem, DM--DE
interaction~(e.g.,~\cite{interaction,chamaleon}) was also considered,
yielding an energy transfer between the dark components, so allowing a
(quasi)--parallel scaling of DM and DE from a fairly high redshift
until the present.  While laboratory data set no significant
constraint on DM--DE interaction strength, parametrized by $\beta$
(see below), recent works placed constraints on possible couplings, by
using SNIa data~\cite{amendolasn} or the redshift evolution of the
Hubble parameter, $H$~\cite{hz}. Accordingly, $\beta >
0.12$--0.15~\cite{dualaxion,diporto} is hardly consistent with
observations.

Unfortunately, such a low coupling level no longer eases the
coincidence problem~\cite{failures}, but, once the genie has come out
from the lamp, it is hard to put it back inside. The point is whether
low values of $\beta$, as allowed by current data, can interfere with
future data analysis. In particular, when we allow for non--zero
$\beta$, how do errors on other parameters behave?

In this work we tried to answer this question by using a Fisher matrix
technique. We considered two different models, set by similar values
of cosmological parameters, without and with coupling. In the latter
case, we took $\beta = 0.1~.$ Starting from these models, we evaluated
the expected errors on cosmological parameters, as obtained when data
concern just CMB anisotropy and polarization or include tomographic
weak lensing (WL).

As a matter of fact, in coupled models, the time evolution of the dark
components is non--standard. If such models are considered in a
Newtonian approximation, it is as though DM particles had a
$\phi$--dependent mass. Also for quite low $\beta$'s, this anomalous
scaling leaves an imprint on both the expansion history of the
Universe, and the growth of (matter) fluctuations, at the linear and
non--linear levels (e.g.~\cite{mattergrow}).

However, any detected evolution of $H$ can be reproduced through a
suitable redshift dependence of DE density $\rho_{de}$ and state
parameter $w_\phi$, when $\phi$ approaches $m_p$ (the Planck mass).  A
risk is that, if matter and dark energy are coupled, fitting
observations leads to an estimate of a phantom equation of state
($w_\phi < -1$), even if $w_\phi > -1$ at all
redshifts~\cite{wphantom}.

In principle, this risk can be excluded if the redshift dependence of
the growth factor $G(z)$ is also tested, through the increase in number
and concentration of bound systems. Data providing information both on
$H(z)$ and $G(z)$ are therefore able to discriminate between coupled
and uncoupled models. Experiments, or combinations of experiments,
probing $H(z)$ and $G(z)$ are then needed.

CMB data, used to constrain coupling~\cite{dualaxion,couplconstr},
place only upper limits on $\beta$. The analysis of Ly--$\alpha$ and
the matter power spectrum of the 2dF and SDSS surveys~\cite{diporto}
does not lead to great improvements. At the available sensitivity
level, such data systems provide just weighted integrals of $H(z)$ and
$G(z)$, which remain consistent with a rather wide set of options.

On the contrary, gravitational lensing, alone or in combination with
CMB data, was already shown to be a powerful tool for the analysis of
DE. WL tomography probes the power spectrum $P(k)$ at different
redshifts and is thus well suited to constrain $G(z)~.$

In this work we aim to put these conceptual points on a more
quantitative basis and to deepen the case of coupling, by performing a
Fisher analysis of future WL surveys and CMB experiments.

The outline of this work is as follow. In Sec.~\ref{sec:def} we review
the basic properties and definitions of dDE models and WL, in
Sec.~\ref{sec:fore} we show the results of the Fisher analysis, in
Sec.~\ref{sec:discu} we discuss them and in Sec.~\ref{sec:summ} we
summarize our findings and draw our conclusions.

\section{Models and definitions}
\label{sec:def}

\subsection{Interacting Dark Energy}
\label{sec:de}
We consider a cosmological model where the DE field $\phi$ interacts
with the cold DM component. The model requires the specification of
the potential $V(\phi)$ and the function $f(\phi)$ characterizing the
coupling. The equation of motion for $\phi$ then reads
\begin{equation}
\label{eq1}
\ddot{\phi} +3H \dot{\phi} = -V^{eff}_{,\phi}~~~~{\rm with} \qquad
V^{eff} = V + \rho_c~~.
\end{equation}
Here dots denote ordinary time differentiation, $H(a) = \dot{a}/a$ and
$\rho_c$ is DM energy density. In turn, its evolution is governed by
\begin{equation}
\dot{\rho}_c +(3H +C\dot{\phi}) \rho_c = 0 ~,\qquad {\rm with} \qquad
C(\phi) = {d\,{\rm log}(f) \over d\phi}~.
\end{equation}
This equation can be integrated and gives:
\begin{equation}
\rho_c (a) = \rho_{ c, 0} a^{-3} f(\phi)~.
\label{eq:rho}
\end{equation}
For $f = 1$ eqs.~(\ref{eq1}), (\ref{eq:rho}) return ordinary dDE
equations. The equations for the other components remain unchanged. In
a generic coupled model, then, the ratio between the energy densities
of cold DM and baryons is not fixed, but evolves in time according to
$f(\phi)$.

However, it is always possible to define an effective DE component of
density
\begin{equation}
\rho_{de}^{eff}(a) = \rho_c a^{-3} [f(\phi)-1]
+\rho_{de} (a)~.
\end{equation}
In general, $\rho_{de}^{eff}(a)$ is not guaranteed to be positive and
detecting $\rho_{de}^{eff}(a) <0$ would be a clear indication that our
description of the dark sector is not adequate. Lacking such clear
giveaway, however, experiments probing $H(z)$ can hardly discriminate
between DE--DM interaction and an {\it ad hoc} DE component. If its
effective state parameter $w^{eff}~ (=p^{eff}/\rho^{eff})\, $ is $<
-1$, data may appear to require phantom DE (see~\cite{das06} for
discussion).

\begin{figure}
\begin{center}
\includegraphics*[width=10cm]{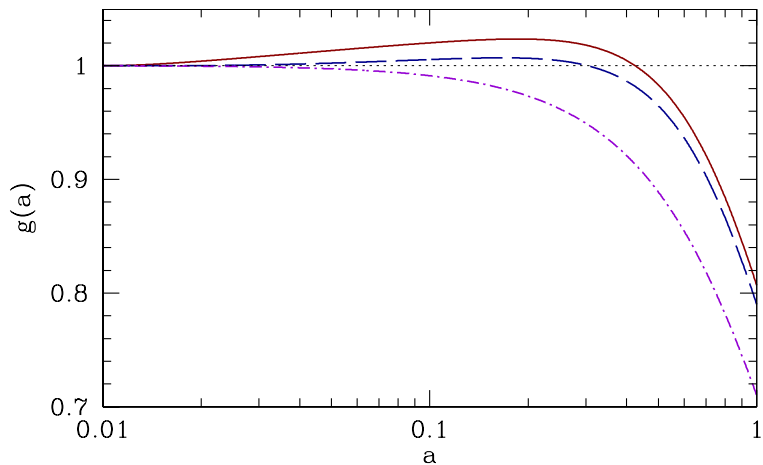}
\end{center}
\caption{\label{fig:lingrowth} The growth suppression rate $g(a)
\equiv \delta (a) /a$ for CDM (solid line) and baryons (dashed line) in
a coupled model. Curves refer to a SUGRA potential with exponential
coupling. It can be noticed that the cold dark matter evolution
rapidly diverges from the standard cold dark matter solution, $g(a) =
1$ (dotted line), even well into the matter dominated era. For
comparison, we also plot $g(a)$ for the same SUGRA model with coupling
turned off (dot--dashed line).}
\end{figure}

Coupling affects also fluctuation growth. In the Newtonian limit, {\it
i.e.} well below the horizon, and neglecting the contribution of
radiation, baryons and DM fluctuations grow according to the equations
(e.g.~\cite{mattergrow,mainini07}):
\begin{eqnarray}
\begin{array}{l}
\ddot{\delta}_{\rm b} +2H\dot{\delta}_{\rm b} = 4\pi G (\rho_{\rm b}
\delta_{\rm b} + \rho_c \delta_c)\\
\ddot{\delta}_c +2H\dot{\delta}_c = 4\pi G \left[ \rho_{\rm b}
\delta_{\rm b} + (1 +{4\over3} \beta^2 )\rho_c\delta_c\right],
\end{array}
\label{eq:delta}
\end{eqnarray}
where we defined $\beta$ so that
\begin{equation}
C(\phi) = 4 \sqrt{\frac{\pi}{3}}\frac{\beta(\phi)}{m_p},
\end{equation}
while $m_p = G^{-1/2}$. Therefore, baryons and DM perturbations grow
at different rates and, even soon after recombination, a growing mode
$ \delta \propto a$ no longer exists (see Fig.~\ref{fig:lingrowth}),
leading to a bias between baryon and DM perturbations. Analytical
models of spherical collapse have shown that this differential growth
results in a baryon--DM segregation, with baryons occupying the outer
regions of collapsed objects~\cite{mainini05}.

Eqs~(\ref{eq:delta}) also show that the growth equations explicitly
depend on both $H$ and $\rho_c$; therefore, if we measure the growth
from data, a possible anomalous scaling can no longer be masked though
an {\it ad--hoc} definition of an effective DE density. It is then
licit to conclude that experiments probing the rate of growth of
fluctuations are in principle well suited to test coupling between the
dark components.

In this work we are interested to combining WL and CMB data,
which cannot be accurately described using current parametrisations
\cite{amendolawl}. Therefore, we follow a more conventional approach and
choose the functional forms
\begin{equation}
V(\phi) = {\Lambda^{4 + \alpha} \over \phi^\alpha} \exp \left( 4\pi
{\phi^2 \over \rm m^2_{PL}}\right)~, \qquad \qquad f(\phi) = \exp \left( \beta
\sqrt{8\pi \over 3} {\phi_0 -\phi \over \rm m_{\rm PL}} \right)
\end{equation}
The SUGRA \cite{sugra} potential $V(\phi)$ depends on the slope
$\alpha$ and the energy scale $\Lambda$. Fixing DE density today and
$\Lambda $ ($\alpha$), however, determines a unique value of $\alpha$
($\Lambda$). CMB, SNIa and deep sample data yield $\Lambda \lesssim
10^3$GeV~\cite{lambdalim}, in the absence of coupling. Here we focus
on the SUGRA potential as it naturally arises in the context of
Supergravity Theories and is an example of tracking potential
characterized by a rapid time variation of the equation of state, when
DE becomes dominant. Therefore, assuming a constant $w$ for this class
of potential may lead to misleading results. Different choices for the
potential are clearly possible.

The coupling function $f(\phi)$ depends on $\beta$, and $\phi_0$ is
the field value today. In this work we assume a constant $\beta \ge 0$
(see however~\cite{das06} for a different approach); data place the
upper limit $\beta \lesssim 0.12-0.15$~\cite{dualaxion,diporto}. For
reasonable values of the cosmological parameters and of $\Lambda$, we
expect coupling effects not to be relevant for $\beta \lesssim 0.01$,
so that the dynamically interesting values for the coupling lies in
the range $0.01 < \beta < 0.10$.

\subsection{Weak lensing}
Among the cosmological probes allowing the analysis of the nature of
DE, the cosmological WL has been earning a fundamental role (see
\cite{Bartelmann:1999yn,Refregier:2003ct,Van
Waerbeke:2003uq,Schneider:2005ka} for a thorough review). In fact,
next generation WL surveys will cover a significant fraction of the
sky and observe galaxies at deeper redshift, making WL a powerful tool
to study the properties of the Universe. Furthermore, WL tomography
will allow to significantly increase the cosmological information that
can be recovered from such surveys (see \cite{Amara:2006kp} and
\cite{Hu:1999ek,Huterer:2001yu,Hu:2003pt} for details about lensing
tomography).

The power spectrum for the WL convergence between the $i$th-- and
$j$th-- redshift bin is given by:
\begin{eqnarray}
  P_{(ij)}(\ell)&=&\left(\frac{H_0}{c}\right)^4
  \int \frac{c\,\rmd z}{H(z)} W_i(z) W_j(z)
  P_{\rm NL}\left(\frac{\ell}{r(0,z)},z\right).
\end{eqnarray}
Here $P_{\rm NL}(\ell/r(0,z),z)$ is the non--linear matter power
spectrum, at the redshift $z$ and wave number $k=\ell/r(0,z)$; the
quantity $r(z,z_s)$ is the radial comoving distance between $z$ and
$z_s$,
\begin{eqnarray}
\label{rzzs}
  r(z,z_s)&=&\int_z^{z_s} \frac{c\,\rmd z'}{H(z')},
\end{eqnarray}
while
\begin{eqnarray}
  W_i(z)&=&\frac3{2}\Omega_{m\,0}F_i(z)(1+z)
\end{eqnarray}
is the window function which weights different redshift bins according
to the factor $F_i(z)$,
\begin{eqnarray}
  F_i(z)&=&\int_{Z_i} \rmd z_s \frac{n_i(z_s)r(z,z_s)}{r(0,z_s)},
\end{eqnarray}
being $Z_i$ the i--th redshift bin. The function $n_i(z)$ is the
normalized redshift distribution of the source galaxies falling in the
$i$-th photometric redshift bin, per unit solid angle,
\begin{eqnarray}
  n_i(z)&=&D_i(z)\left[\int_0^{\infty} \rmd z'\,D_i(z')\right]^{-1},
\end{eqnarray}
where
\begin{eqnarray}
  D_i(z)&=&\int_{z_{ph}^{(i)}}^{z_{ph}^{(i+1)}} \rmd z_{ph} \,n(z)\,
  p(z_{ph}|z).
  \label{eq:13}
\end{eqnarray}
The last expression codifies the effect of errors in photometric
redshifts of source galaxies \cite{Ma:2005rc}. In fact, due to the
large number of observed galaxies with future surveys, one needs to
refer to their photometric redshifts, even if the determination of
these redshifts may be not so accurate as the spectroscopic ones. In
the model given by \eref{eq:13}, the mapping between the photometric
$z_{ph}$ and the spectroscopic redshift $z$ is obtained convolving the
overall galaxy distribution per unit solid angle, $n(z)$, with a
probability distribution $p(z_{ph}|z)$ in $z_{ph}$ at a given $z$. We
choose a Gaussian function at each redshift for the distribution of
photometric redshifts, i.e.
\begin{eqnarray}
\label{eq:photoz}
  p(z_{ph}|z)=\frac1{\sqrt{2\pi} \sigma_z}\exp{
    \left[-\frac{(z-z_{ph})^2}{2\sigma_z^2}\right]}
\end{eqnarray}
while the overall distribution of source galaxies is chosen to have
the parametrized form:
\begin{equation}
  n(z) = \frac{d^2N}{dz\,d\Omega} = \frac{B}{z_0\,\Gamma\left(
    \frac{A+1}{B}\right)}
  \left(\frac{z}{z_0}\right)^A \exp{\left[-(z/z_0)^B\right]},
  \label{eq:distro}
\end{equation}
where $A$, $B$ and $z_0$ are the parameters. One can easily check that
equation \eref{eq:13} turns into:
\begin{eqnarray}
  D_i(z)=\frac1{2}n(z)\left[{\rm erf}(x_{i+1})-{\rm
  erf}(x_i)\right],
\end{eqnarray}
with $x_i\equiv (z_{ph}^{(i)}-z)/\sqrt 2\sigma_z$ and ${\rm erf}(x)$
the error function.

The cosmic shear power spectrum will receive a shot-noise contribution
from the random intrinsic ellipticities of source galaxies and
measurement error \cite{Kaiser:1996tp}. Thus, the observed power
spectrum between redshift bins $i$ and $j$ can be expressed as:
\begin{eqnarray}
  P^{\rm obs}_{(ij)}(\ell)=P_{(ij)}(\ell)+\delta_{ij}
  \frac{\sigma_\epsilon^2}{\bar{n}_i}
\end{eqnarray}
where $\sigma_\epsilon$ is the rms shear due to intrinsic ellipticity
and measurement noise (we assume $\sigma_\epsilon\simeq 0.22$
\cite{Amara:2006kp}) and
\begin{eqnarray}
  \bar{n}_i = \left[\frac{n_g}{amin^{-2}}\right]
  \left(\frac1{60}\frac{\pi}{180}\right)^{-2}\hat{n}_i
\end{eqnarray}
is the average number density of galaxies per steradians in the $i$-th
redshift bin, $n_g$ being the number of galaxies per square arcminute
and $\hat{n}_i$ the fraction of sources belonging to the bin.

The covariance between the power spectra $P^{\rm obs}_{(ij)}(\ell)$
and $P^{\rm obs}_{(mn)}(\ell ')$ is approximately given by
\begin{eqnarray}
  \fl{\rm Cov}\left[P^{\rm obs}_{(ij)}(\ell),
        \,P^{\rm obs}_{(mn)}(\ell ')\right]=
      \frac{\delta_{\ell\ell '}}{(2\ell+1)\Delta\ell f_{sky}}
      \left[P^{\rm obs}_{(im)}(\ell)\,P^{\rm obs}_{(jn)}(\ell ')+
    P^{\rm obs}_{(in)}(\ell)\,P^{\rm obs}_{(mj)}(\ell ')\right]
      \label{covar}
\end{eqnarray}
where $f_{sky}$ is the sky fraction covered by the survey and
$\Delta\ell$ is the bin width centred at $\ell$. The above expression
assumes that the power spectrum in each multipole bin is very flat, in
order to replace the value of the spectrum evaluated at the bin center
with the average of spectrum over each bin (see \ref{Append:A} for the
complete expression). In addition, we have not included the
non-Gaussian term, due to the contribution of the shear
trispectrum~\cite{Cooray:2000ry,Takada:2003ef}.

\subsection{Fisher's formalism}
The Fisher matrix formalism
\cite{Sivia:1996,Fisher:1935,Tegmark:1996bz} provides lower limits to
the error bars of the cosmological parameters one wishes to
measure. The basic tool in Fisher's method is the likelihood function,
yielding the probability that a model gives the set of data
$\textit{\textbf{x}}$.

Suppose we want to test an hypothesis, {\it i.e.} a cosmological model
set by $M$ parameters $\btheta=(\theta_1,\,\theta_2,\,\ldots\,
\theta_M)$. The likelihood function
$L(\textit{\textbf{x}}|\btheta)=\exp [-\cal
L(\textit{\textbf{x}}|\btheta)]$ is often a complicated function of
$\btheta$; the value $\hat{\btheta}$ corresponding to the peak of $L$
defines the {\it maximum likelihood estimator} which, in the limit of
large data sets, becomes the {\it best unbiased estimator} of the
actual parameter set. Thus, the likelihood can be Taylor expanded to
second order (the first non--vanishing term) around $\hat{\btheta}$,
being so approximated with a multivariate Gaussian distribution
\begin{eqnarray}
  L(\textit{\textbf{x}}|\btheta)
  \propto
  \exp{\left(-\frac1{2}\Delta\btheta^T{\bf C}(\btheta)^{-1}
    \Delta\btheta\right)}~;
  \label{gaussian}
\end{eqnarray}
here
\begin{eqnarray}
  {\bf C}(\btheta)^{-1}=\left.\frac{\partial^2 {\cal
    L}(\textit{\textbf{x}}|\btheta)} {\partial \theta_{\alpha}
    \partial \theta_{\beta}} \right|_{\btheta=\hat{\btheta}}
       \label{covarmtx}
\end{eqnarray}
is a positive semi-definite non-singular matrix, dubbed {\it
covariance matrix} of the $\theta_{\alpha}$. We remind that
\Eref{gaussian} holds just in a sufficiently small neighborhood around
the maximum $\hat{\btheta}$. In turn, the {\it Fisher information
matrix} reads
\begin{eqnarray}
  {\bf F}_{\alpha\beta}(\btheta)=\left\langle\frac{\partial{\cal L}
    (\textit{\textbf{x}}|\btheta)}{\partial \theta_{\alpha}}
  \frac{\partial{\cal L}
    (\textit{\textbf{x}}|\btheta)}{\partial \theta_{\beta}}\right\rangle
  _{\btheta=\hat{\btheta}}=
  \left\langle-\frac{\partial^2{\cal L}(\textit{\textbf{x}}|\btheta)}
  {\partial \theta_{\alpha} \partial \theta_{\beta}}\right\rangle
  _{\btheta=\hat{\btheta}}~;
  \label{fisher}
\end{eqnarray}
the average $\langle\ldots\rangle=\int {\cal
L}(\textit{\textbf{x}}|\btheta) \ldots d^N\textit{\textbf{x}}$ is
taken over all possible data realizations, given the model
parameters. The feature making Fisher's formalism so significant is
the Cram\'{e}r-Rao theorem. It states that the parameter variance
about any unbiased estimator value owns a lower bound:
$\Delta\theta_{\alpha}\geq \sqrt{({\bf F}^{-1})_{\alpha\alpha}}$, if
the other parameters are estimated from the data as well,
$\Delta\theta_{\alpha}\geq 1/\sqrt{{\bf F}_{\alpha\alpha}}$, if all
the other parameters are known.  Therefore,the Fisher information
matrix components are the expectation values of ${\bf
C}^{-1}(\hat{\btheta})$. Accordingly, the inverse of the Fisher matrix
is an estimate of the covariance matrix of the parameters ${\bf
C}(\btheta)\thickapprox {\bf F}^{-1}$.

A convenient way to re--write the Fisher matrix is computing the
derivatives of the likelihood function using the following chain rule
\cite{Oh:1998sr}:
\begin{eqnarray}
  \left.\frac{\partial {\cal L}(\textit{\textbf{x}}|\btheta)}
     {\partial \theta_{\alpha}}\right|_{\btheta=\hat{\btheta}}
     =\sum_{\ell}
     \left.\frac{\partial {\cal L}(\textit{\textbf{x}}|\btheta)}
     {\partial x_{\ell}}\right|_
     {\textit{\textbf{x}}=\textit{\textbf{x}}(\hat{\btheta})}
     \left.\frac{\partial x_\ell}{\partial \theta_\alpha}
     \right|_{\btheta=\hat{\btheta}}
\end{eqnarray}
Thus, the Fisher matrix \eref{fisher} can be expressed as:
\begin{eqnarray}
  {\bf F}_{\alpha\beta}(\btheta)
  &=&\sum_{\ell\ell'}
     \frac{\partial x_\ell}{\partial \theta_\alpha}
     \frac{\partial x_{\ell'}}{\partial \theta_\beta}
     \left\langle-\frac{\partial^2 {\cal L}(\textit{\textbf{x}}|\btheta)}
     {\partial x_{\ell} \partial x_{\ell'}}\right\rangle_
     {\textit{\textbf{x}}=\textit{\textbf{x}}(\hat{\btheta})}\\
  &=&\sum_{\ell\ell'}
     \frac{\partial x_\ell}{\partial \theta_\alpha}
     {\bf F}_{\ell\ell'}(\btheta)
     \frac{\partial x_{\ell'}}{\partial \theta_\beta}\\
  &\thickapprox&\sum_{\ell\ell'}
     \frac{\partial x_\ell}{\partial \theta_\alpha}
     {\bf C}^{-1}_{\ell\ell'}(\btheta)
     \frac{\partial x_{\ell'}}{\partial \theta_\beta},
\end{eqnarray}
where ${\bf F}_{\ell\ell'}$ and ${\bf C}_{\ell\ell'}$, respectively,
are the Fisher and the covariance matrix for the observables
$\textit{\textbf{x}}$. The region in the $M$--dimensional space of the
parameters, defined by $Q(\theta,\hat{\theta})=\Delta\theta_{\alpha}^T
{\bf F}_{\alpha\beta}\Delta\theta_\beta=K^2$, is a hyper--ellipsoid of
constant probability density for the function \eref{gaussian}.
Marginalizing over the other parameters, one can project this
ellipsoid in the two--parameter subspace, yielding a two--dimensional
ellipse. The analytical expression for the projected ellipse for the
two parameters $\theta_\alpha$ and $\theta_\beta$ is given by
\cite{Matsubara:2002cs}:
\begin{eqnarray}
  \left(
  \begin{array}{cc}
    \Delta\theta_{\alpha}&\Delta\theta_{\beta}
  \end{array}
  \right)
  \left[
  \begin{array}{cc}
    ({\bf F}^{-1})_{\alpha\alpha}&({\bf F}^{-1})_{\alpha\beta}\\
    ({\bf F}^{-1})_{\alpha\beta}&({\bf F}^{-1})_{\beta\beta}
  \end{array}
  \right]^{-1}
  \left(
  \begin{array}{c}
    \Delta\theta_{\alpha}\\\Delta\theta_{\beta}
  \end{array}
  \right)=\Delta\chi^2(N=2,\sigma)
\end{eqnarray}
This can be interpreted as an estimate of the confidence region within
a given confidence level $\sigma$ for the two parameters
$\theta_\alpha$ and $\theta_\beta$.

\section{Forecasts for Future Experiments}
\label{sec:fore}
We present here the results of the Fisher analysis of future
experiments, considering both WL and CMB measurements. For
definiteness we assume a fiducial WL survey with characteristics
similar to those of the recently proposed DUNE project
\cite{Refregier:2006vt}. We assume a redshift distribution of the form
\eref{eq:distro} with $A=2$, $B=1.5$ and $z_0\simeq z_{\it m}/1.412$,
corresponding to a median redshift of the survey $z_{\it m}=0.9$ (see
\cite{Amara:2006kp}), and a mean surface density of galaxies $n_g =
35\,{\rm arcmin}^{-2}$. The full survey, covering half of the sky
($f_{sky}=0.5$), is divided into $N=5$ redshift bins, with
$p(z_{ph}|z)$ given by equation~\eref{eq:photoz} and
$\sigma_z(z)=0.05(1+z)$.

We consider lensing multipoles up to $\ell_{\rm max} = 20000$, since
we find that results do not depend significantly on larger $\ell$.
However, one should bear in mind that when considering scales $\ell
\gg 1000$ there could be some non--linear and baryonic effects on the
matter power spectrum, and so on the WL spectrum
\cite{Rudd:2007zx}. These effects, not yet well understood, could be
important for forecasts. However, in this work we suppose these
effects to be negligible.

For CMB data, we consider an ideal experiment with characteristics
based on the 143GHz PLANCK channel: angular resolution $\theta_{\rm
  fwhm} = 7.1'$ and sensitivity $\sigma_{\rm T} = 42 \mu{\rm K\,
  arcmin}$, $\sigma_{\rm P} = 80 \mu{\rm K\, arcmin}$.

The cosmological model we consider is characterized by 7 parameters
with fiducial values:
\begin{eqnarray}
\nonumber
\vec{\theta} = \{~& \omega_{\rm b} & =  (0.045\cdot0.7^2),~\omega_{\rm m} =
(0.30\cdot0.7^2),~\Omega_{\rm m} = 0.30,~n_{\rm s} = 1.00, \\
\nonumber
& \sigma_8 & = 0.8,~\Lambda_{\rm DE} = 5\cdot10^{-3} {\rm GeV},~\beta = 0.1~\}.
\end{eqnarray}
Here $\Omega_{\rm m}$ represents the current total (CDM + baryons)
matter density in units of the critical density; $\omega_{\rm b}
\equiv \Omega_{\rm b}~h^2$ and $\omega_{\rm m} \equiv \Omega_{\rm
m}~h^2$ are the physical baryons and total matter densities,
respectively; $n_{\rm s}$ is the slope of the primordial power--law
spectral index of density fluctuations; $\sigma_8$ is the rms mass
fluctuation in spheres of 8$h^{-1}$~Mpc radius while $\Lambda_{\rm
DE}$ and $\beta$ were defined in Sec.~\ref{sec:de}. Let us notice that
the class of DE models considered here reduces to $\Lambda$CDM for
$(\Lambda_{\rm DE})^4 \simeq 10^{-47} {\rm GeV}^4$ and $\beta = 0$.
Moreover, the fiducial values of DE parameters $\Lambda_{de}=
5\cdot10^{-3} {\rm GeV}$ and $\beta = 0.1$ are chosen in order to
reproduce at $z=0$ an effective equation of state which mimics the
case of $\Lambda$CDM model, $w = -0.95$. Finally, when dealing with
CMB data, we also need to fix the value of the optical depth to
reionization, $\tau = 0.10$.

We compute the CMB anisotropies (temperature and polarisation) power
spectra and the transfer functions, used to calculate linear matter
power spectrum, using a modified version of CAMB~\cite{camb}. To evaluate the
non--linear matter power spectrum, $P_{NL}$, we employ the
prescription by Smith {\it et al.} \cite{Smith:2002dz}. This is only
tested for model with a cosmological constant; as we are concerned
here with Fisher matrix estimates assume that the results
of~\cite{Smith:2002dz} can be extended to coupled models simply by
taking into account the non--standard scaling of $\rho_c$
(eq.~\ref{eq:rho}). Numerical derivatives were evaluated considering
a $5\%$ stepsize, except for $\Lambda_{\rm DE}$, where we adopted a
$5\%$ stepsize on $\lambda \equiv Log_{10} (\Lambda_{\rm DE} /{\rm
GeV})$.

\subsection{CMB measurements}

\begin{table}[t!]
\caption{\label{tab:errors} Estimated errors on model parameters.}
\begin{indented}
\item[]
\begin{tabular}{@{}lllllll}
\br
&\multicolumn{2}{c}{CMB}&\multicolumn{2}{c}{WL}&\multicolumn{2}{c}{WL+CMB}\\

& \tiny{SUGRA} & \tiny{SUGRA}& \tiny{SUGRA}& \tiny{SUGRA}
&\tiny{SUGRA}& \tiny{SUGRA} \\

& \tiny{$\beta =0.$} & \tiny{$\beta =0.1$} & \tiny{$\beta =0.$} &
\tiny{$\beta =0.1$} & \tiny{$\beta =0.$} & \tiny{$\beta =0.1$}\\
\mr 100$*\omega_{\rm b}$ &0.016 &0.019 &0.5   &0.9    &0.011  &0.012 \\
$\omega_{\rm m}$         &0.002 &0.006 &0.016 &0.03   &0.0004 &0.0005\\
$\Omega_{\rm m}$         &0.05  &0.12  &0.002 &0.0014 &0.0011 &0.0014\\ 
$n_s$                   &0.004 &0.005 &0.012 &0.018  &0.0014 &0.0021\\
$\sigma_8$              &0.07  &0.13  &0.0026&0.0029 &0.0017 &0.0016\\
 $\lambda$              &7.2   & 9.5  &0.89  &1.1    &0.28   &0.28 \\ 
$\beta$                 & --   &0.04  & --   &0.018  & --    &0.0016\\
$\tau$                  &0.005 &0.006 & --   & --    & --    & -- \\   \br
\end{tabular}
\end{indented}
\end{table}

Table~\ref{tab:errors} lists the estimated errors on the various
parameters considered. For each data set, we compare forecasts for the
target model with results for a SUGRA model with the same values of
the relevant parameters. The table clearly shows that a PLANCK--like
experiment is able to provide a measurement of a direct DE--DM
interaction at 68\% confidence level, even for moderate values of the
coupling strength $\beta$. However, we expect that at 90\% confidence
level data will still be compatible with $\beta = 0$.

In any case, allowing for a direct interaction strongly degrades the
experimental sensitivity on the parameters characterizing the matter
density and the normalization of the primordial spectrum of density
fluctuations. Errors on these quantities increase by a significant
amount.

\begin{figure}
\begin{center}
\includegraphics*[width=10cm]{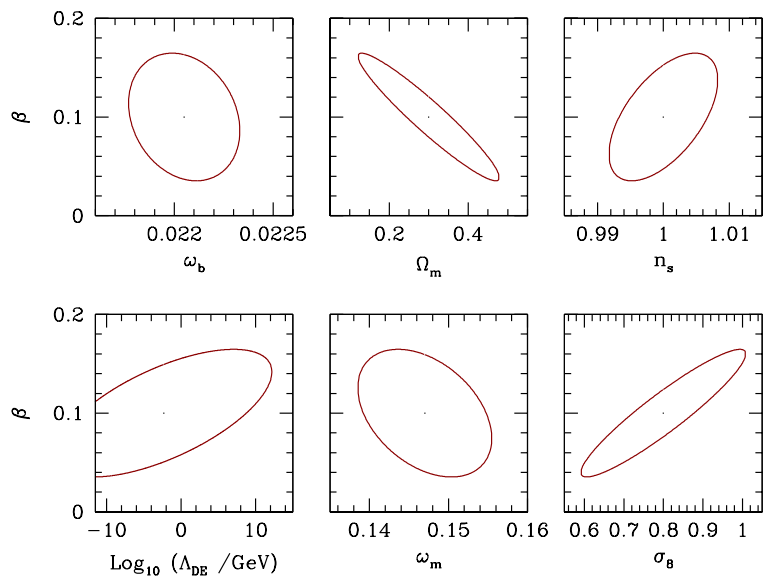}
\end{center}
\caption{\label{fig:cmb} Forecasts of joint 1--$\sigma$ confidence
regions on the coupling parameter $\beta = 0.1$ and selected
parameters, for a PLANCK--like experiments, after full marginalization
over the remaining parameters.}
\end{figure}

Figure~\ref{fig:cmb} shows the joint $68\%$ confidence regions between
$\beta$ and each of the other parameters, except for $\tau$,
considering only CMB data. In each plot we marginalized over the
parameters not shown. $\beta$ is strongly correlated with most
parameters considered here, with the exception of $\omega_{\rm b}$
(and $\tau$), thus introducing additional degeneracies in actual data
analysis.

A detailed characterization of these degeneracies would require a
different approach than that followed here (e.g. Monte Carlo Markov
Chains simulations). We just point out that they can be understood
recalling that the heights of the acoustic peaks of CMB spectra are
sensitive to the total matter density and to baryon/dark matter ratio
at last scattering. In coupled models, these quantities are not
univocally determined by their present day value, but strongly depend
on $\beta$. In addition, the total growth between the last scattering
epoch and today is strongly sensitive to $\beta$, resulting in a clear
degeneracy between $\beta$ and $\sigma_8$.

Let us notice that, as we are concerned with a combination of CMB and
WL data, the parameter set is not optimized for CMB experiments. Using
a different parametrisation can alter error estimates and/or the
degeneracies between the various parameters. In particular, CMB data
are better described in terms of the angle subtended by the acoustic
horizon at recombination, $\theta$, and the amplitude of the
primordial spectrum of density fluctuations $A_s$, rather than in
terms of $\Omega_m$ and $\sigma_8$. This results in large errors on
the latter parameters, when CMB data alone are considered. Adopting a
set of parameters better suited to the analysis of CMB data results in
slightly lower error estimates overall, but the effects of coupling
are largely unchanged.

\subsection{Weak Lensing}

\begin{figure}
\begin{center}
\includegraphics*[width=10cm]{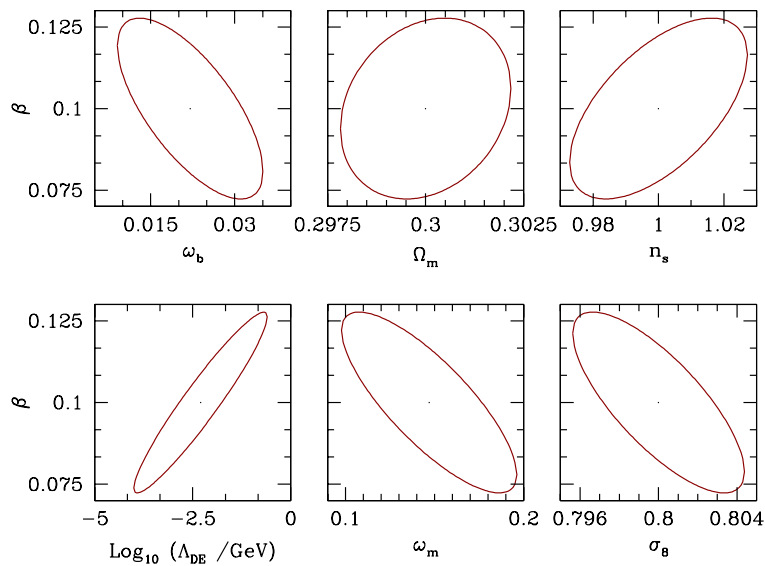}
\end{center}
\caption{\label{fig:wl} Forecasts of joint 1--$\sigma$ confidence
regions on the coupling parameter $\beta = 0.1$ and selected
parameters, for a DUNE--like experiments, after full marginalization
over the remaining parameters. Notice the change of scales with
respect to \fref{fig:cmb}}
\end{figure}
\begin{figure}[t]
    \begin{center}
        \includegraphics[width=6.5cm]{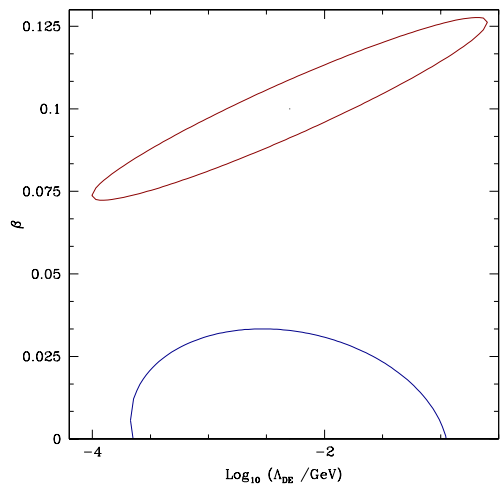}
    \end{center}
  \caption{\label{fig:compar} Comparison between the 1--$\sigma$
    confidence regions of a coupled SUGRA model with $\beta = 0.1$
    (red) and a non--coupled SUGRA model with $\beta = 0$ (blue) for a
    WL experiment. It is clearly possible to distinguish between the
    two models.}
\end{figure}
\begin{figure}[t]
    \begin{center}
        \includegraphics[width=6.5cm]{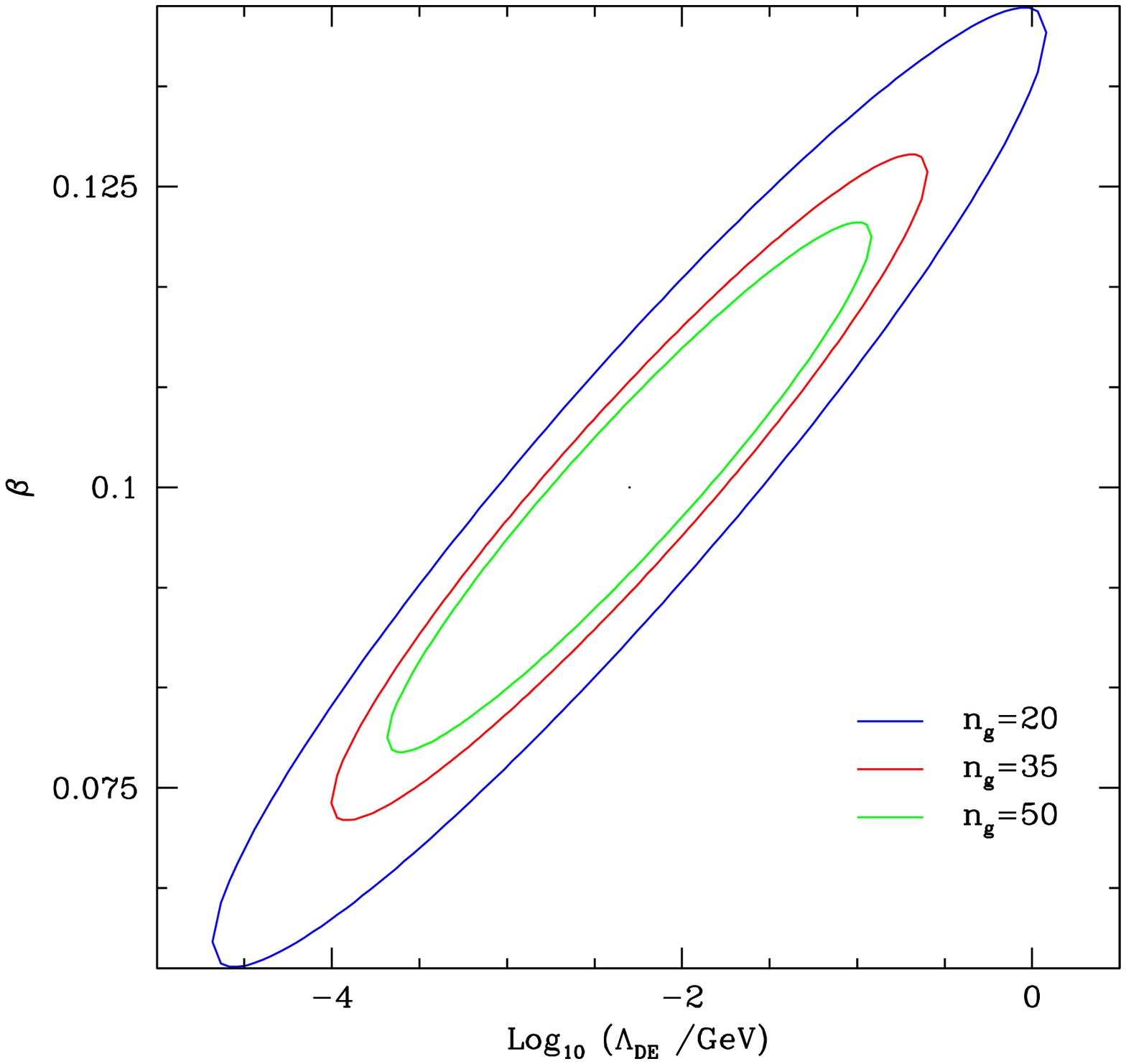}
        \includegraphics[width=6.5cm]{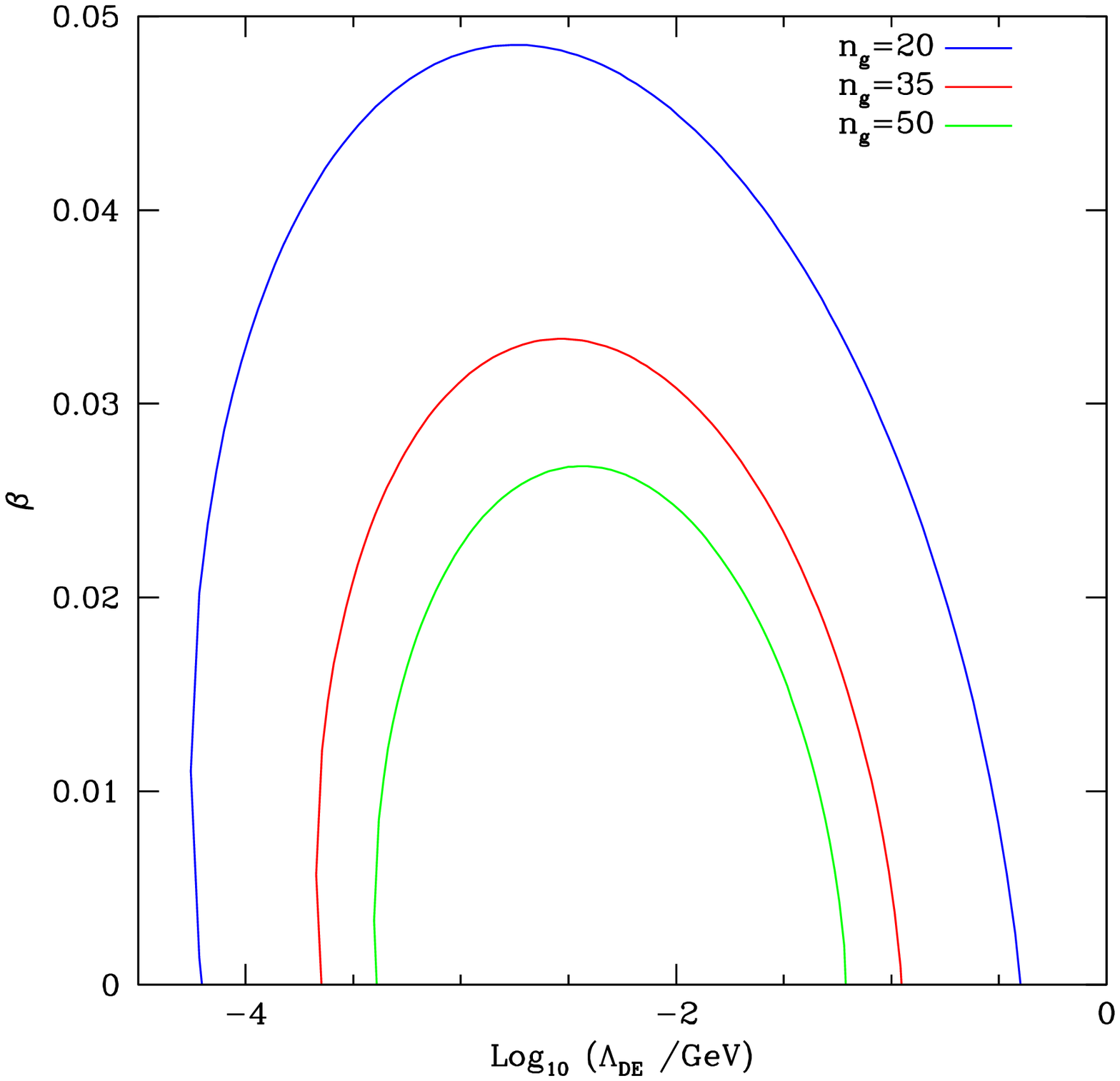}
    \end{center}
  \caption{\label{fig:b+l} Joint 1--$\sigma$ confidence regions on
    $\beta$ and $\Lambda_{\rm DE}$ after marginalization over the
    remaining parameters, for different number of galaxies $n_g$. On
    the left SUGRA coupled model with $\beta = 0.1$, on the right
    SUGRA coupled model with $\beta = 0$ and $\beta$ derivatives
    calculated only on one side, for positive values of the
    parameter.}
\end{figure}
\begin{figure}[t]
    \begin{center}
        \includegraphics[width=6.5cm]{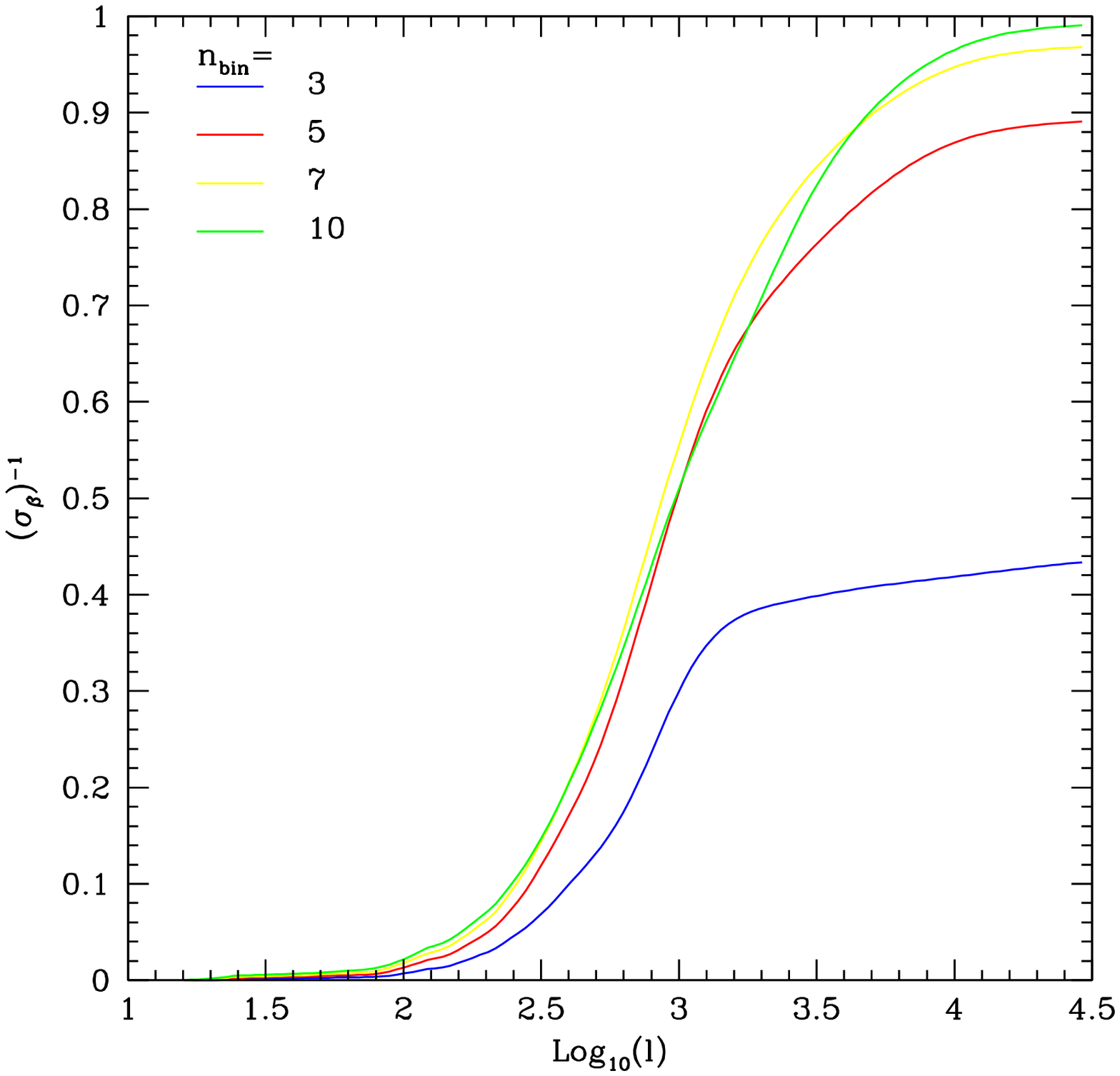}
        \includegraphics[width=6.5cm]{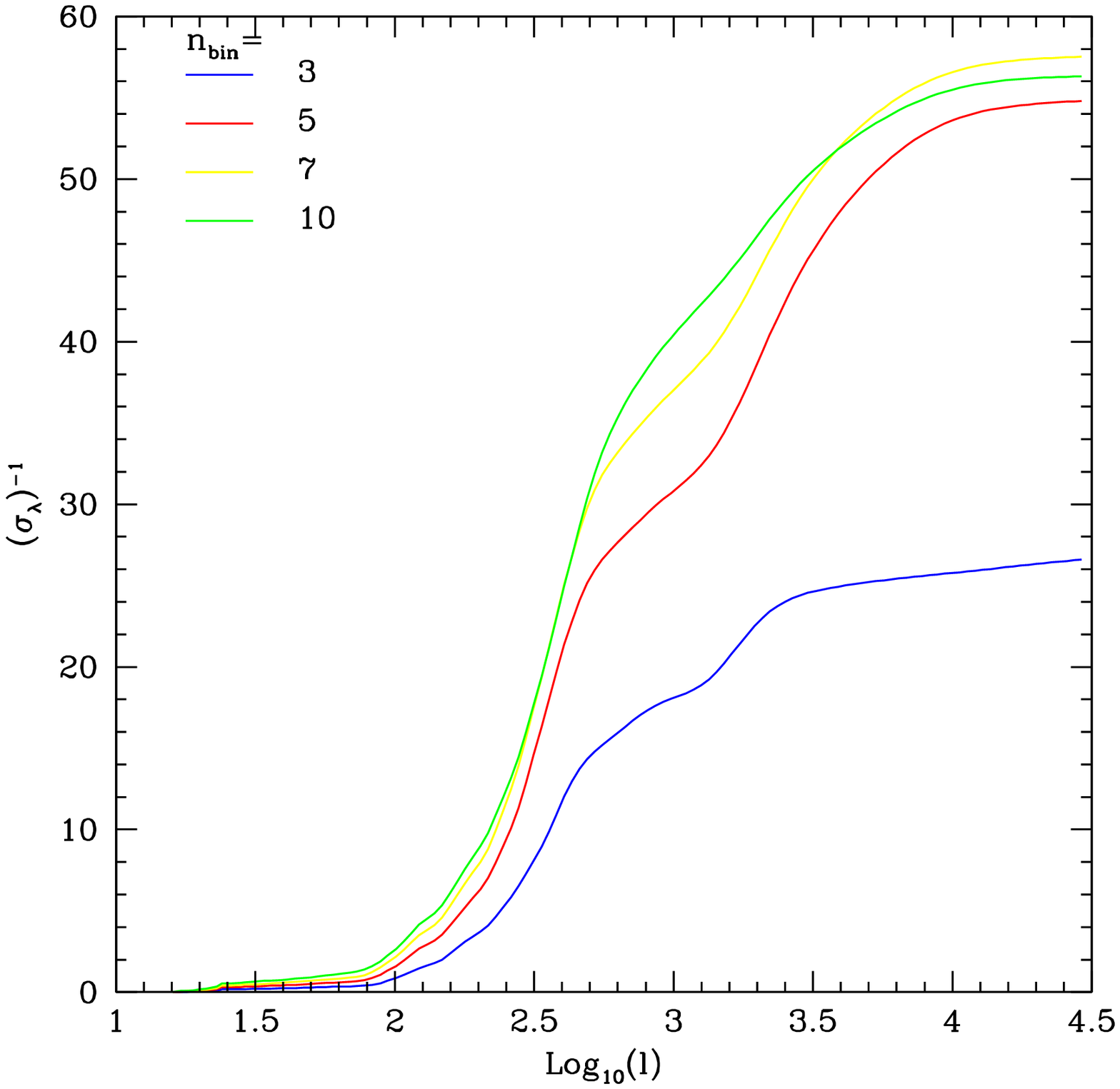}
    \end{center}
  \caption{\label{fig:inverr} Inverse error as a function of the
  maximum multipole for the WL survey, for different numbers of
  redshift bins. Left panes show results for $\beta$, right panel
  refers to $\lambda$.}

\end{figure}

\Fref{fig:wl} is analogous to \fref{fig:cmb} for our target weak
lensing survey. Together with Table~\ref{tab:errors}, these results
show the great potential of WL surveys in constraining interacting DE
models. Marginalized errors on $\beta$ and $\lambda$ are of the order
of $\sigma(\beta) \simeq 0.02$ and $\sigma(\lambda) \simeq 1 $; these
figures represent a factor of 2, or more, improvement over Planck
estimates. WL data alone can clearly distinguish the target model from
a non--coupled model or a cosmological constant even at the
3$\sigma$-level, {\it viceversa} assuming a reference SUGRA model with
$\beta = 0$, we can expect to put an upper limit $\beta \lesssim
0.03$, at the same confidence level (see~\fref{fig:compar}).

As expected, WL surveys perform significantly better than CMB
experiments also with respect to parameters specifying the current
matter density, $\Omega_m$ and $\sigma_8$. Moreover, constraints on
these parameters are not significantly affected by the coupling
degrees of freedom. Errors on the remaining parameters, instead,
increase by a factor of $\sim 2$. Finally, we consider a combination
of CMB and WL data. CMB and WL probe very different epoch of the
Universe and are sensitive to different combination of cosmological
parameters. Considering both CMB and WL data allows to constrain the
DE parameters with a few percent accuracy, and significantly reduces
the degeneracies introduced by DE coupling. In this case, the errors
on the cosmological parameters are very similar in both models
considered, with the exception of $n_s$.

Next we considered how our results depend on the characteristic
assumed for the target survey. In \fref{fig:b+l} the impact of mean
surface density of galaxies on the determination of $\beta$ and
$\lambda$. With $n_g = 25\,{\rm arcmin}^{-2}$ constraints on $\beta$
degrades by $\sim 50\%$, while $n_g = 50\,{\rm arcmin}^{-2}$ gives
only a marginal improvement on expected errors; constraints on
$\lambda$ are similarly affected. However, even in the worst case
considered here, next generations WL survey will provide an
improvement over the information that we are likely to obtain from
PLANCK data.

Lastly, we consider the dependence of our results on the number of
bins and the multipoles range considered. In \fref{fig:inverr} we plot
the inverse of the expected variance on $\beta$ and $\lambda$ as a
function of the maximum multipole considered in the analysis and for
different number of bins. With 3 redshift bins, the precision on both
parameters depends mostly on multipole up to a few thousands; smaller
scales do not provide a significant contribution. Dividing the survey
in 5 bins strongly improves the constraints on both parameters and
allows to exploit information from multipoles up to $\sim 10000$. For
a DUNE--like survey, a further increase in the number of bins does not
lead to significant improvements on the constrains on coupled models
parameters.

It must be outlined that these results assume that the theoretical
framework used to predict the matter power spectrum on intermediate
($1 h{\rm Mpc^{-1}} \lesssim k \lesssim 20 h{\rm Mpc^{-1}}$) and small
($k > 20 h{\rm Mpc^{-1}}$) scales can accurately account for the
effects of baryons on non--linear structures.  In general, the fitting
formulas used to predict the non--linear power spectrum are calibrated
using dissipationless N--body simulations and, therefore, do not
properly describe baryonic structures.  While baryons make up $\simeq
15-20\%$ of the matter in the Universe and on large scales are
expected to trace the DM field, their distribution inside halos is
significantly different from DM. In turn, this alters the shape of the
non--linear power spectrum on the corresponding scales, and the
possibility of extracting precision constraints from $P_{\rm NL}(k)$
hinges on our capability of accurately modeling baryon
physics~\cite{Rudd:2007zx}. However, simulations do not yet have the
accuracy required for precision constraints and the problem is even
more serious for the coupled models considered in this work. Modelling
non linear stages though spherical growth, Mainini
\cite{Mainini:2005fe} shew that baryons and DM will be however
differently distributed, even independently of the onset of gas
dynamics. N--body simulations of cDE models were performed
\cite{Maccio:2003yk}, by using a Ratra--Peebles \cite{Ratra:1987rm}
potential; hydro simulations, instead, were never produced.  Should
accurate prediction be still unavailable for the analysis of a
DUNE--like experiment, a more conservative cutoff of $l\simeq 1000$
would be required. \Fref{fig:inverr} shows that in this case the
expected errors on $\Lambda$ and $\beta$ would increase by a factor
$\sim 2$.

\section{Discussion}
\label{sec:discu}

All previous analysis shows that, even if we admit quite a little
DM--DE coupling, we open a Pandora's box, leading to a severe
degradation in our capacity to deduce cosmological parameters from a
given set of measures.

As a matter of fact, coupling destroys our trust that the period
between the recombination and the start of DE relevance is under
control. If coupling is absent, during such period SCDM is a fair
approximation. Let us then remind what happens to the growth factor,
as soon as coupling is onset: Figure \ref{fig:lingrowth} shows that:
(i) deviations from SCDM are significant already when $a \sim 0.02$;
(ii) they are then different for DM and baryons; (iii) they work in
the opposite direction, in respect to the effects of a DE components.

As far as the growth factor is concerned, a tiny coupling is able to
overwhelm a huge DE amount, with compensation occurring for $a \sim
0.3$--0.4$\, $, however keeping always $g(a)$ at values greater by
$\sim 10$--15$\, \% \, $.  Altogether, growth is faster in coupled
models. Hence, if we do not include the information that coupling is
zero in the fit, we can find an agreement between data and a wider
range of DE amounts.

\begin{figure}
  \begin{center}
    \includegraphics[width=8.5cm]{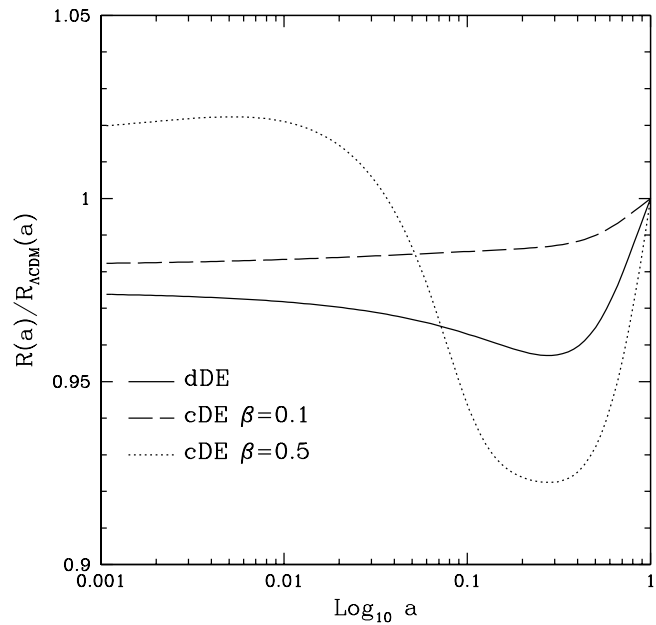}
    \caption{\label{fig:distar} Scale dependence of comoving distances
in coupled or uncoupled SUGRA cosmologies, compared with $\Lambda$CDM.
The case $\beta = 0.5$, corresponding to a rather strong DM--DE
coupling is also shown.}
  \end{center}
\end{figure}

Similar points can be made for the comoving radial distance $R(z)
\equiv r(0,z)$ (see eq.~\ref{rzzs}). In Figure \ref{fig:distar} we
compare comoving distances for $\Lambda$CDM with various cosmologies.
The Figure shows that dDE, in the absence of other parameter shifts,
sets the Last Scattering Band (LSB) closer to the observer. Once
again, a mild coupling acts in the opposite direction and tends to
re-set the LSB at the distance it had in $\Lambda CDM$. In the Figure
we consider the behavior of distances also for a rather strong
coupling, $\beta = 0.5$. Then the distance behavior is different in
the period when DE density can be neglected, in respect to the epoch
when DE and DM have similar densities. The key point, however, is that
the LSB becomes then farther from the observer. When fitting CMB data
to such models, in order to compensate such effect, the value of $H_o$
tends to be increased. Strong coupling therefore yields a large Hubble
parameter estimate.

\begin{figure}[t]
   \centering 
   \includegraphics[width=9.3truecm]{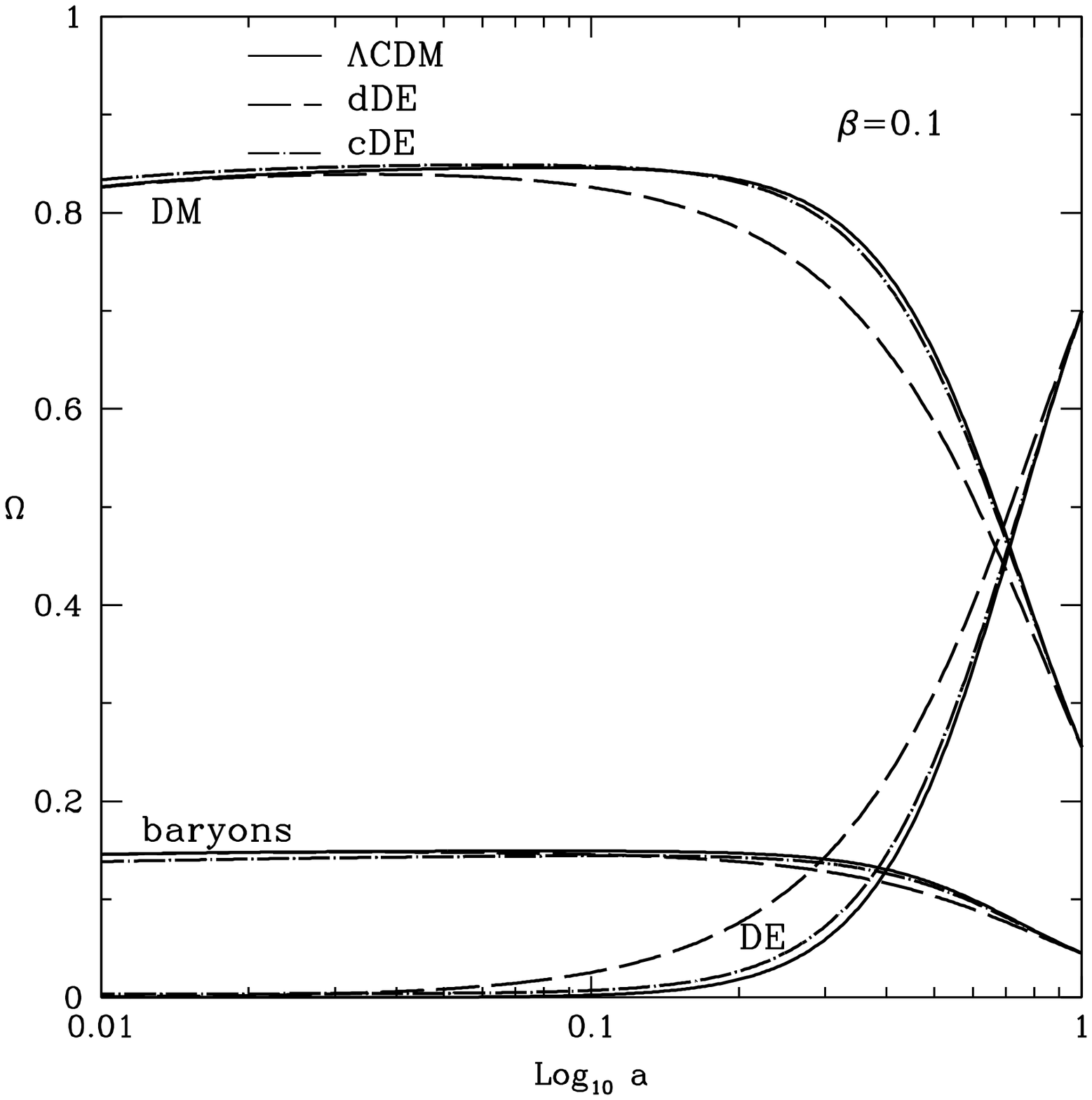}
   \vskip .6truecm
   \includegraphics[width=9.3truecm]{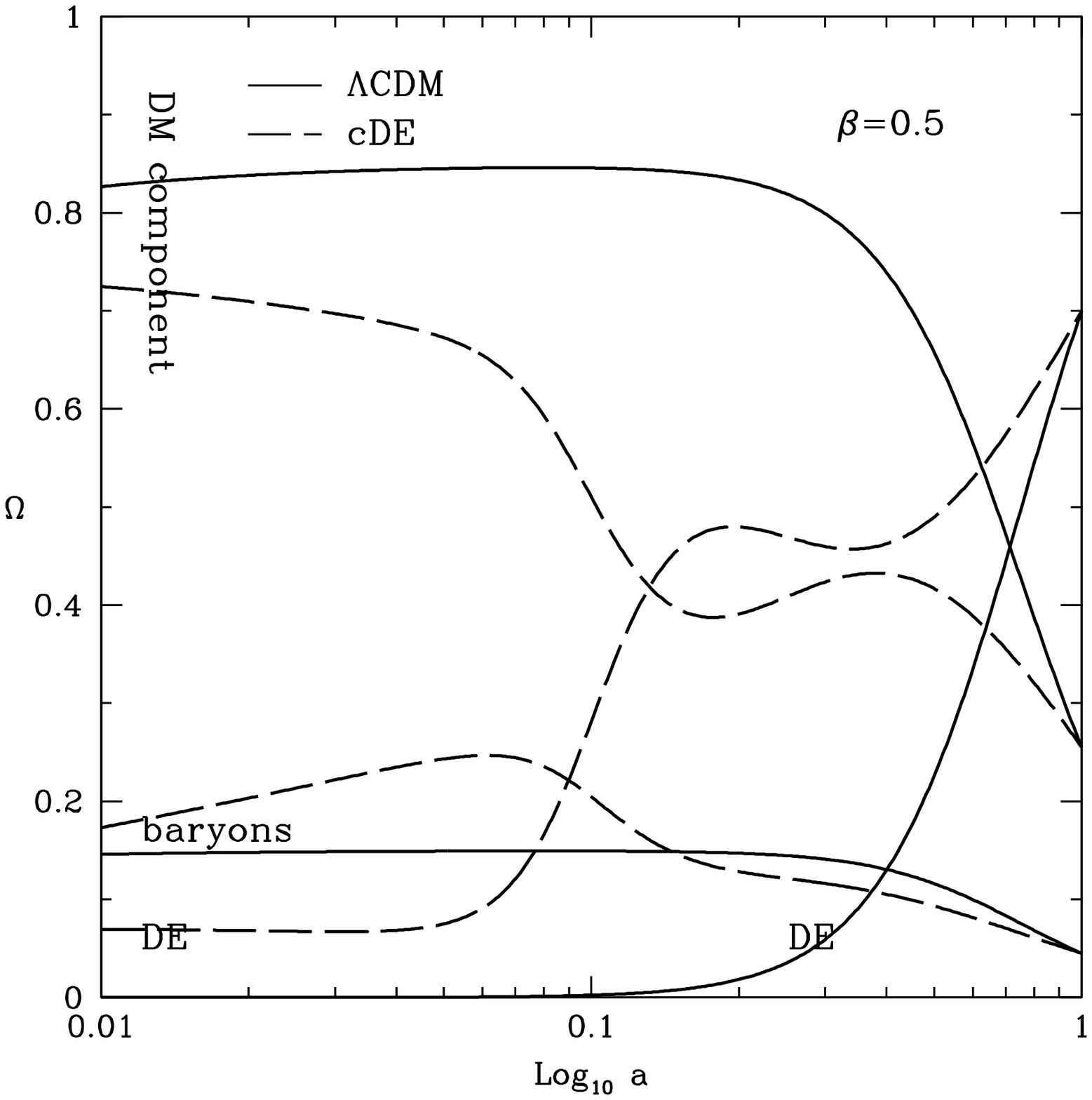}
   \caption{\label{fig:omega} Scale dependence of the density
parameters of the various components in $\Lambda$CDM. compared with
other models.  In the upper panel uncoupled and weakly coupled SUGRA
models are considered. For the sake of comparison, in the lower panel
we also show the effects of choosing a stronger coupling.  }
\end{figure}

Figures \ref{fig:omega} finally show the scale dependence of the
density parameters in the different models.  Once again, when DE is
mildly coupled, a behavior more similar to $\Lambda CDM$ is
recovered. On the contrary, when considering a greater coupling
strength, we see that DE and DM keep similar densities up to a fairly
large redshift. This was indeed the initial motivation of cDE
cosmologies.

Altogether, these Figures indicate that adding a small coupling
reduces the effects of the very passage from $\Lambda CDM$ to dDE;
owing to the excellent fit that $\Lambda CDM$ cosmologies have with
data, this tells us that only highly refined CMB data will be able to
test the possibility that a mild DM--DE coupling exists.

On the contrary, a stronger coupling, although easing the coincidence
problem, displaces several observables in a unacceptable way.

\section{Summary and Conclusions}
\label{sec:summ}

\begin{figure}[t]
\begin{center}
\includegraphics*[width=10cm]{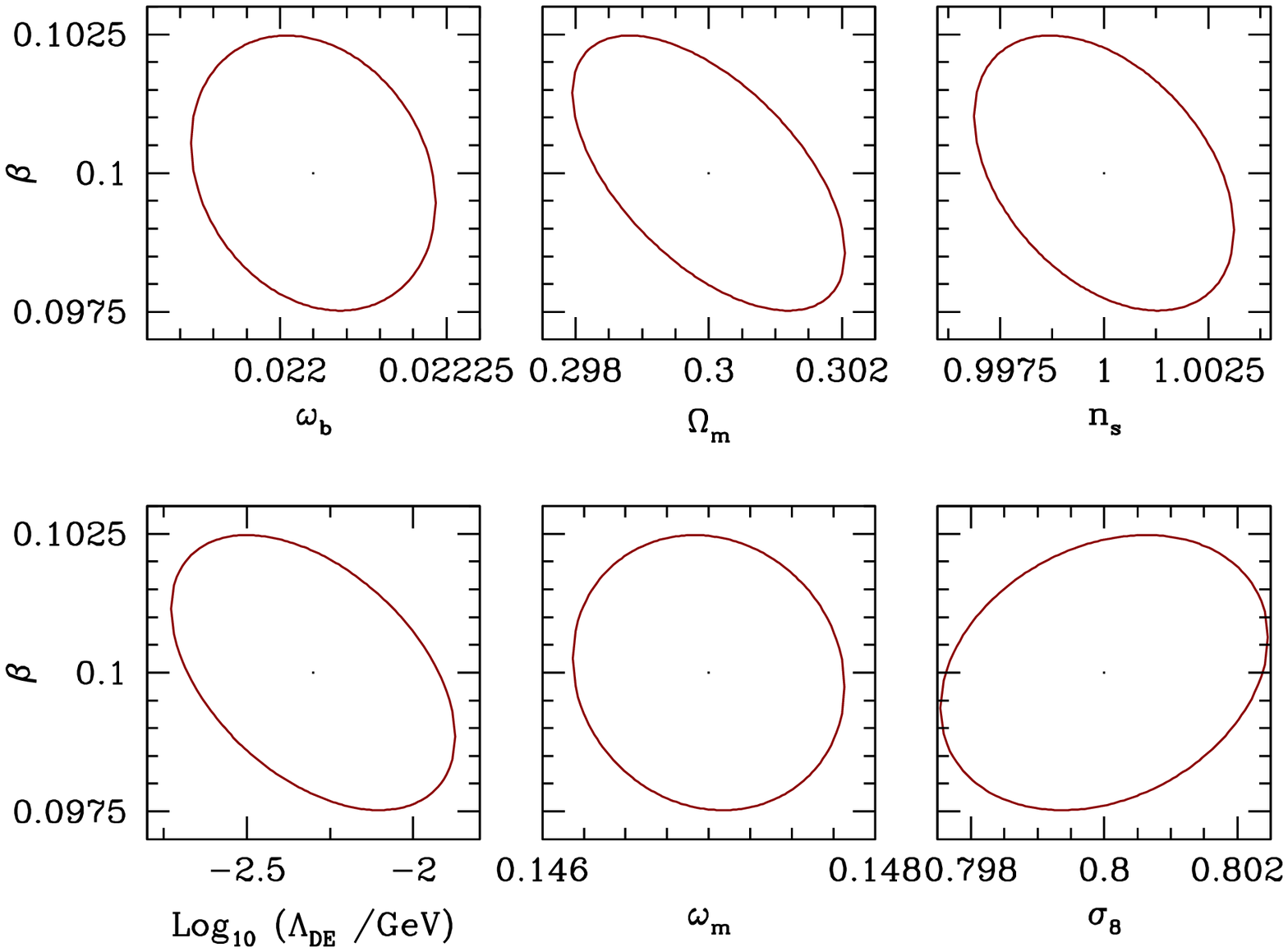}
\end{center}
\caption{\label{fig:wlcmb} Forecasts of joint 1--$\sigma$ confidence
regions on the coupling parameter $\beta = 0.1$ and selected
parameters, for a combination of a PLANCK--like and a DUNE--like
experiment, after full marginalization over the remaining
parameters.
}
\end{figure}

Future WL surveys will certainly put more stringent constraints on
cosmological parameters and will be crucial to break quite a few
degeneracies between parameters.

Within this context, in this paper we focus on coupled DE models with
a twofold aim. Detecting a signal of DM--DE coupling would be
certainly decisive to fix the nature of the dark components.
Henceforth determining the level of sensitivity needed to appreciate
such an effect is crucial in setting the appeal of forthcoming
projects. There is however a complementary aspect which deserves much
attention. In order to convert raw data into physical information, a
set of parameters, spanning a variety of models, is to be fixed; a
bias on parameter selection, however, can lead to an optimistic
estimate of the confidence level for the best fitting model, far from
reality.

In this paper, we focused on this kind of danger, when we open the
option of DM--DE coupling. Even if such coupling is absent or quite
weak, we showed that just considering its possibility may widen the
error bars for a number of parameters, also apparently unrelated to
the coupling itself. In a sense, when a new degree of freedom is
opened, such an effect is natural and expected. Coupling, however, has
really a major impact, affecting different parameters for the
different observables considered here; moreover, its impact is
drastically reduced when we work out parameter values by using
simultaneously both observables.

Before outlining our main conclusions it is however worth reminding a
technical limit we had to face. While CMB predictions depend on linear
spectra, the WL spectrum is limited to its non--linear shape. For the
purpose of the present analysis, we assumed that prescription for
$\Lambda$CDM models \cite{Smith:2002dz} can be trivially extrapolated
to coupled models, so enabling us to estimate the non linear spectrum
once the linear spectrum is known. Let us however remark that the
shift was estimated from quite a wide set of $\Lambda$CDM
simulations. Simulations of cosmological models with state parameters
$w \neq 1$, although performed by several authors, are still not so
extensively studied as $\Lambda$CDM; let alone coupled DE simulations:
in this case the only available simulations are due to
\cite{Maccio:2003yk} and deal with a potential $V(\phi)$ different
from SUGRA.

However, differences between prescriptions for $\Lambda$CDM and dDE
are small \cite{Ma:2006zk} and it seems however clear that model
differences can most affect the rate of evolution of halo
concentration, slightly shifting the scale where non--linearity
effects become significant. The use of more precise prescriptions can
therefore only cause minor variation on the estimated errors and,
although welcome, such simulations are not expected to interfere
substantially with our conclusions.

Our estimates were based on assuming that a photometric survey is
available, with $\sigma_z(z)=0.05(1+z)$ and a median redshift
$z_m=0.9$, covering half of the sky ($f_{sky}=0.5$). These features
are similar to the recently proposed DUNE experiment \cite{
Refregier:2006vt}. We also compared and combined results from WL with
the constraints expected for an ideal Planck--like experiment. The
basic results of our calculations are quoted in table~\ref{tab:errors}
which is one of the main results of this work.

A first set of conclusions concerns an ideal CMB experiment considered
by itself. In this case, introducing coupling degrees of freedom is
crucial for the error estimates on some of parameters, also apparently
unrelated to coupling. In particular, while the error on $\omega_b$
keeps $\sim$1\%, the errors on $\omega_m$ and $\sigma_8$ increase from
1.3\% to 3.9\% and from 9\% to 16\%, respectively.

From a physical point of view, the option opened by coupling is that
expansion rate and fluctuation growth, from the last scattering band
to the observer's site, is non--standard ({\it e.g.}, the
proportionality law $\rho_m \propto a^{-3}$ could be mildly
violated). Although CMB data themselves set stringent limits on such
deviations, this widens the volume of the parameter space consistent
with a given data set; in particular, it increases the likelihood of
values of $H_0$ that would otherwise be negligible and, because of the
intercorrelation amongst parameters, this reflects immediately on
$\omega_m$ and $\Omega_m$ estimates.

Similar effects occur in WL experiments, although involving different
parameters. Such experiments are a direct test of $\Omega_m$, whose
estimated error is reduced by a factor $> 30$ in respect to a CMB
experiment. When the coupling option is opened, the error on
$\Omega_m$ does not increase; on the contrary, it becomes easier to
attribute raw data uncertainties to other parameters and, although
marginally, the error on $\Omega_m$ becomes somehow smaller.

A completely new situation occurs if both CMB and WL measurements are
simultaneously used. In this case, the opening of the coupling option
causes just a marginal increase of the errors on most parameters.
This is a clear indication of the complementarity of the CMB and WL
measurements, as described in \cite{Eisenstein:1998hr} and one of our
conclusion is that the combination of these observables, besides of
providing parameter values independent of $\beta$, can set a
(nearly--)final word on the coupling option.

As a matter of fact, by comparing \fref{fig:cmb}, \fref{fig:wl} and
\fref{fig:wlcmb}, we see that, when joining CMB and WL results, the
degeneracies, between $\beta$ and $\sigma_8$, as well as between
$\beta$ and $\Omega_m$, disappear. Breaking degeneracies is the main
aim when different observables are simultaneously considered. We see
that, from this point of view, the efficiency of using both CMB and WL
measures can be hardly overestimated.

Let us then focus on the case of the spectral index of scalar
fluctuations. When both CMB and WL data are used to constrain $n_s$, a
sharp reduction of errors occurs. No surprise that CMB data, by
themselves, reflecting the state of the universe before the onset of
non linear processes, were more efficient to constrain $n_s$, with or
without coupling, than WL measures. Joining together the two
observables, we then see errors to decrease from 0.4\% and 1.2\% down
to 0.16\%, in the uncoupled case; in the presence of coupling we have
a similar behavior, with errors passing from 0.5\% and 1.8\% down to
0.25\%. The error level achieved, in both cases, is exceptional, even
for precision cosmology, and clearly suggests to relieve the
constraint of a single $n_s$ value, so inspecting its possible scale
dependence, with realistic possibilities to find a direct insight into
the nature of the inflationary potential.

In conclusion, future WL surveys could really allow a significant step
forward in the comprehension of the dark cosmic side; we can adfirm
that, when they will be available, the endeavour to put the genie back
inside the lamp will approach a full success.

\ack LPLC is supported by NASA grant NNX07AH59G and JPL--Planck
subcontract no. 1290790. We acknowledge the use of the CAMB package.
LPLC and GLV thank S. Bonometto for insightful discussion and help with the
preparation of the manuscript.

\appendix
\section{Convergence power spectrum covariance}
\label{Append:A}

In order to determine the convergence power spectrum covariance, one
can introduce the so called ``flat-sky'' approximation and treat the
sky as flat, replacing spherical harmonic sums with Fourier transforms
(FT). Of course, this approximation is acceptable just for small
angular scales We also consider the tomographic case and use Greek
letters as superscripts to denote quantities belonging to different
redshift bins. The FT of the convergence field can be defined as:
\begin{eqnarray}
  \kappa^{\alpha}\left({\bf l}\right)=\int \rmd^2\btheta \,\kappa^{\alpha}
  (\btheta)\exp{(-i\btheta\cdot{\bf l})},
\end{eqnarray}
while the convergence power spectrum and trispectrum are, respectively:
\begin{eqnarray}\label{eq:A.2}
  \langle \kappa^{\alpha}\left( \mathbf{l}_{1} \right)
  \kappa^{\beta} \left( \mathbf{l}_{2} \right) \rangle  = 
  \left( 2\pi \right)^2 \delta_D \left( \mathbf{l}_{1}+\mathbf{l}_{2} \right)
  P^{\alpha\beta}_{l_{1}},
\end{eqnarray}
\begin{eqnarray}\label{eq:A.3}
  \langle \kappa^{\alpha}\left( \mathbf{l}_{1} \right)\ldots
  \kappa^{\delta} \left( \mathbf{l}_{4} \right) \rangle_{c}  = 
  \left( 2\pi \right)^2 \delta_D \left( \mathbf{l}_{1}+\ldots
  +\mathbf{l}_{4} \right)
  T^{\alpha\beta\gamma\delta}\left(\mathbf{l}_1,\mathbf{l}_{2},\mathbf{l}_{3}, 
  \mathbf{l}_{4} \right);
\end{eqnarray}
here $\delta_D$ is the Dirac function. The value of the lensing power 
spectrum in correspondence of a multipole $l$ can be estimated as the mean 
over a multipole bin of width $\Delta l$, centered at $l$:
\begin{eqnarray}
  \mathcal{P}_{\it{l}}^{\alpha\beta}
  & = & \frac{1}A\int_{s_l} \frac{\rmd^2\mathbf{l}_1}{A_{l}}
   \kappa^{\alpha} \left( \mathbf{l}_1 \right)
  \kappa^{\beta} \left( -\mathbf{l}_1 \right),
\end{eqnarray}
where $A_{l}=\int_{s_l}\rmd^2\mathbf{l}\cong 2\pi l\,\Delta l$ is the
area of the shell of width $\Delta l$ corresponding to $l$, while
$A=4\pi f_{sky}$ is the area of the survey. Quite in the same way, for
the trispectrum we have:
\begin{eqnarray}
  \mathcal{T}_{\it{ll'}}^{\alpha\beta\gamma\delta}  = 
  \int_{s_l}\frac{\rmd^2\mathbf{l}_1}{A_{l}}
  \int_{s_{l'}}\frac{\rmd^2\mathbf{l}_2}{A_{l'}}
  T^{\alpha\beta\gamma\delta}
  \left( \mathbf{l}_{1}, -\mathbf{l}_{1},\mathbf{l}_{2},
  -\mathbf{l}_{2} \right).
\end{eqnarray}
Let us then consider the following expression:
\begin{eqnarray}\label{eq:A.6}
  \langle \mathcal{P}^{\alpha\beta}_{l}
  \mathcal{P}^{\gamma\delta}_{l'} \rangle  = 
  \frac{1}{A^2}\int_{s_l}\frac{\rmd^2\mathbf{l}_1}{A_{l}} 
  \int_{s_{l'}}\frac{\rmd^2\mathbf{l}_2}{A_{l'}}
  \langle 
  \kappa^{\alpha}\left( \mathbf{l}_{1} \right)
  \kappa^{\beta} \left( -\mathbf{l}_{1} \right)
  \kappa^{\gamma}\left( \mathbf{l}_{2} \right)
  \kappa^{\delta} \left( -\mathbf{l}_{2} \right) \rangle.
\end{eqnarray}
The 4-point function at the r.h.s. can be decomposed in its connected
parts:
\begin{eqnarray}
  \fl
  \langle \kappa^{\alpha}\left( \mathbf{l}_{1} \right)
  \kappa^{\beta} \left( -\mathbf{l}_{1} \right)
  \kappa^{\gamma}\left( \mathbf{l}_{2} \right)
  \kappa^{\delta} \left( -\mathbf{l}_{2} \right) \rangle =
  \langle \kappa^{\alpha}\left( \mathbf{l}_{1} \right)
  \kappa^{\beta} \left( -\mathbf{l}_{1} \right)
  \kappa^{\gamma}\left( \mathbf{l}_{2} \right)
  \kappa^{\delta} \left( -\mathbf{l}_{2} \right) \rangle_c+\nonumber\\
  +\langle \kappa^{\alpha}\left( \mathbf{l}_{1} \right)
  \kappa^{\beta} \left( -\mathbf{l}_{1} \right)\rangle
  \langle\kappa^{\gamma}\left( \mathbf{l}_{2} \right)
  \kappa^{\delta} \left( -\mathbf{l}_{2} \right) \rangle+\nonumber\\
  \fl\quad\quad\quad
  +\langle \kappa^{\alpha}\left( \mathbf{l}_{1} \right)
  \kappa^{\gamma} \left( \mathbf{l}_{2} \right)\rangle
  \langle\kappa^{\beta}\left( -\mathbf{l}_{1} \right)
  \kappa^{\delta} \left( -\mathbf{l}_{2} \right) \rangle+
  \langle \kappa^{\alpha}\left( \mathbf{l}_{1} \right)
  \kappa^{\delta} \left( -\mathbf{l}_{2} \right)\rangle
  \langle\kappa^{\gamma}\left( \mathbf{l}_{2} \right)
  \kappa^{\beta} \left( -\mathbf{l}_{1} \right) \rangle.
\end{eqnarray}
Replacing their expression in eq. \eref{eq:A.6}, one can easily
recognize the contribution of the trispectrum, using
eq. \eref{eq:A.3}:
\begin{eqnarray}
  \fl
  \langle \mathcal{P}^{\alpha\beta}_{l}
  \mathcal{P}^{\gamma\delta}_{l'} \rangle  = 
  \frac{1}{A^2}\int_{s_l}\frac{\rmd^2\mathbf{l}_1}{A_{l}}
  \int_{s_{l'}}\frac{\rmd^2\mathbf{l}_2}{A_{l'}}
  \left( 2\pi \right)^2 \delta_D \left( \mathbf{0} \right)
  T^{\alpha\beta\gamma\delta}\left(\mathbf{l}_1,-\mathbf{l}_{1},\mathbf{l}_{2},
  -\mathbf{l}_{2} \right)
  +\langle \mathcal{P}^{\alpha\beta}_{l}\rangle
  \langle\mathcal{P}^{\gamma\delta}_{l'} \rangle+\nonumber\\
  +\frac{1}{A^2}\int_{s_l}\frac{\rmd^2\mathbf{l}_1}{A_{l}} 
  \int_{s_{l'}}\frac{\rmd^2\mathbf{l}_2}{A_{l'}}
  \langle \kappa^{\alpha}\left( \mathbf{l}_{1} \right)
  \kappa^{\gamma} \left( \mathbf{l}_{2} \right)\rangle
  \langle\kappa^{\beta}\left( -\mathbf{l}_{1} \right)
  \kappa^{\delta}  \left( -\mathbf{l}_{2} \right) \rangle+\label{eq:A.8}\\
  +\frac{1}{A^2}\int_{s_l}\frac{\rmd^2\mathbf{l}_1}{A_{l}}
  \int_{s_{l'}}\frac{\rmd^2\mathbf{l}_2}{A_{l'}}
  \langle \kappa^{\alpha}\left( \mathbf{l}_{1} \right)
  \kappa^{\delta} \left( -\mathbf{l}_{2} \right)\rangle
  \langle\kappa^{\gamma}\left( \mathbf{l}_{2} \right)
  \kappa^{\beta} \left( -\mathbf{l}_{1} \right) \rangle,\label{eq:A.9}
\end{eqnarray}
where $\left( 2\pi \right)^2 \delta_D \left( \mathbf{0} \right)=A$.
Owing to the definition of covariance,
\begin{eqnarray}
  \rm{Cov}\left[ \mathcal{P}^{\alpha\beta}_{\it{l}},
    \mathcal{P}^{\gamma\delta}_{\it{l'}}\right] \equiv
  \langle \mathcal{P}^{\alpha\beta}_{\it{l}}
  \mathcal{P}^{\gamma\delta}_{\it{l'}} \rangle 
  -\langle \mathcal{P}^{\alpha\beta}_{\it{l}}\rangle
  \langle\mathcal{P}^{\gamma\delta}_{\it{l'}} \rangle,\label{eq:A.10}
\end{eqnarray}
and using \eref{eq:A.2}, one can then argue that the integrals in
\eref{eq:A.8} make sense only if they correspond to the same $l$--bin;
the same can be claimed for \eref{eq:A.9}. This property can be
described introducing a Kronecker delta function $\delta_{ll'}$. Thus,
the expression \eref{eq:A.10} becomes:
\begin{eqnarray}
  \fl
  \rm{Cov}\left[ \mathcal{P}^{\alpha\beta}_{\it{l}},
	\mathcal{P}^{\gamma\delta}_{\it{l'}}\right]  &= &
   \frac{1}{A}\mathcal{T}_{\it{ll'}}^{\alpha\beta\gamma\delta}
  + \delta_{\it{ll'}}\frac{\left( 2\pi \right)^2}{A^2} 
  \int_{s_{\it{l}}}\frac{\rmd^2\mathbf{l}_1}{A_{\it{l}}^2}
  \left[ P_{l_{1}}^{\alpha\gamma}
  \langle\kappa^{\beta}\left( -\mathbf{l}_{1} \right)
  \kappa^{\delta} \left( \mathbf{l}_{1} \right) \rangle
  +P_{l_{1}}^{\alpha\delta} \langle\kappa^{\gamma}
  \left( \mathbf{l}_{1} \right)
  \kappa^{\beta} \left( -\mathbf{l}_{1} \right) \rangle \right]\nonumber\\
   &=& \frac{1}{A}\left\{\mathcal{T}_{\it{ll'}}^{\alpha\beta\gamma\delta} +
    \delta_{\it{ll'}} \frac{\left( 2\pi \right)^2}{A_{s_{\it{l}}}}
  \int_{s_{\it{l}}}\frac{\rmd^2\mathbf{l}_1}{A_{s_{\it{l}}}} 
  \left[ P_{l_{1}}^{\alpha\gamma}P_{l_{1}}^{\beta\delta} 
    + P_{l_{1}}^{\alpha\delta}P_{l_{1}}^{\gamma\beta}\right]\right\}\\
  &\thickapprox&\frac{1}{4\pi\,f_{sky}}
  \mathcal{T}_{\it{ll'}}^{\alpha\beta\gamma\delta}+
  \frac{\delta_{\it{ll'}}}{2l\Delta l f_{sky}}
  \left[ P_{l}^{\alpha\gamma}P_{l}^{\beta\delta} 
    + P_{l}^{\alpha\delta}P_{l}^{\gamma\beta}\right].
\end{eqnarray}
In the last line, we have supposed the lensing power spectrum to be
smooth enough to treat it as a constant within each bin width.

\end{document}